\documentclass{PoS}

\def\NIMA#1#2#3{Nucl. Inst. Methods {\bf A#1} (#2) #3}

\title{Measurement of the FCNC decays $K^\pm\to\pi^\pm l^+ l^-$ by the
NA48/2 experiment at CERN}

\ShortTitle{Measurement of the $K^\pm\to\pi^\pm l^+ l^-$ decays by
NA48/2}

\author{\speaker{Evgueni Goudzovski}\\
        University of Birmingham\\
        E-mail: \email{eg@hep.ph.bham.ac.uk}}

\abstract{A sample of 7253 $K^\pm\to\pi^\pm e^+e^-$ decay candidates
with 1.0\% background contamination has been collected by the NA48/2
experiment at the CERN SPS, which allowed a precise measurement of
the form factor, the branching ratio, and the CP violating
asymmetry of $K^+$ and $K^-$ decay widths was investigated. The results of the $K^\pm\to\pi^\pm e^+e^-$ analysis, as well as the status of the $K^\pm\to\pi^\pm\mu^+\mu^-$ analysis
based in the same data set, are reported.}

\FullConference{2009 KAON International Conference KAON09,\\
         June 09 - 12 2009\\
         Tsukuba, Japan}

\begin{document}

\section*{Introduction}

The FCNC processes $K^\pm\to\pi^\pm l^+l^-$
($l=e,\mu$) are induced at one-loop level in the Standard Model. Their decay rates are dominated by the long-distance
contribution via one-photon exchange, and have been described by the Chiral
Perturbation Theory (ChPT). Several models of the vector form
factor characterizing the dilepton invariant mass spectrum and
the decay rate have been proposed~\cite{da98,fr04,du06}.
The first observation of the $K^+\to\pi^+e^+e^-$ process was made at
CERN more than 30 years ago~\cite{bl75}, followed by BNL E777~\cite{al92} and E865~\cite{ap99} measurements.
The most precise of these, E865, based on a sample of 10300
candidates, allowed a detailed analysis of the
decay form factor and rate, and a test of the next-to-leading order
ChPT calculation~\cite{da98}.

A recent precise measurement of the $K^\pm\to\pi^\pm
e^+e^-$ decay based on the full data set collected in
by the NA48/2 experiment at the CERN SPS and published in 2009~\cite{piee} is reported here. The status and prospects of the $K^\pm\to\pi^\pm\mu^+\mu^-$ analysis based on the same data set are also discussed.

%%%%%%%%%%%%%%%%%%%%%%%%%%%%%%%%%%%%%%%%%%%%%%%%%
\section{The NA48/2 experiment at CERN}

The NA48/2 experiment which took data in 2003--04, designed for charge asymmetry
measurements, uses simultaneous $K^+$ and $K^-$ beams
produced by primary SPS protons impinging on a beryllium
target. Charged particles with momentum $(60\pm3)$ GeV/$c$ are
selected by an achromatic system of four dipole magnets with zero
total deflection, which splits the two beams in the
vertical plane and then recombines them on a common axis. The decay volume housed in a 114 m long cylindrical vacuum tank. Both beams follow the same path in the decay volume: their axes coincide within 1~mm, while the transverse
size of the beams is about 1~cm. With $7\times 10^{11}$ protons
incident on the target per SPS spill of $4.8$~s duration, the
positive (negative) beam flux at the entrance of the decay volume is
$3.8\times 10^7$ ($2.6\times 10^7$) particles per pulse, of which
$5.7\%$ ($4.9\%$) are $K^+$ ($K^-$). The $K^+/K^-$ flux ratio is
1.79.

A description of the NA48 detector and 2003--04 data taking can be found
in~\cite{fa07}.
The decay volume is followed by a magnetic
spectrometer housed in a tank filled with helium at
atmospheric pressure, separated from the vacuum tank by a thin
($0.31\%X_0$) Kevlar window. The spectrometer consists of four
drift chambers (DCHs), two located upstream and two downstream
of a dipole magnet which provides a horizontal
momentum kick of $\Delta p=120~{\rm MeV}/c$ for charged particles. The
nominal spectrometer momentum resolution is $\sigma_p/p = (1.02
\oplus 0.044\cdot p)\%$ ($p$ in GeV/$c$).

The spectrometer is followed by a plastic scintillator
hodoscope (HOD) used to produce fast trigger signals and to provide
precise time measurements of charged particles.
The HOD is followed by a liquid krypton electromagnetic calorimeter
(LKr) used for photon detection and particle identification. It is
an almost homogeneous ionization chamber, $27X_0$ deep, segmented transversally into 13248 cells 2$\times$2 cm$^2$ each, with no longitudinal
segmentation. Detectors located further downstream (hadron calorimeter, muon counter) are not used in the $K^\pm\to\pi^\pm e^+e^-$ analysis.

%%%%%%%%%%%%%%%%%%%%%%%%%%%%%%%%
\boldmath
\section{$K^\pm\to\pi^\pm e^+e^-$ decay analysis}
\unboldmath

\noindent The $K^\pm\to\pi^\pm e^+e^-$ rate is measured relative to $K^\pm\to\pi^\pm\pi^0_D$ normalisation channel
(where $\pi^0_D\to e^+e^-\gamma$ is the Dalitz decay). The
signal and normalisation final states contain
identical sets of charged particles. Thus particle
identification efficiencies, potentially a significant
source of systematic uncertainties, cancel at first order.
Three-track vertices (compatible with
$K^\pm\to\pi^\pm e^+e^-$ and $K^\pm\to\pi^\pm\pi^0_D$ topology) are
reconstructed by extrapolation of track segments from the
spectrometer into the decay volume,
accounting for stray magnetic fields and multiple scattering.

A large part of the selection is common to signal and normalisation modes, and requires a presence of a vertex satisfying the following criteria.\newline
-- Vertex longitudinal position is inside fiducial decay volume:
$Z_{\rm vertex}>Z_{\rm final~collimator}$.\newline
-- The tracks should be
in DCH, HOD and LKr geometric acceptance, and have momenta
in the range $5~{\rm GeV}/c<p<50~{\rm GeV}/c$. Track separations should exceed 2~cm in DCH1 plane to suppress $\gamma$
conversions, and 15~cm in LKr front plane to minimize
effects of shower overlaps.\newline
-- Total charge of the three tracks: $Q=\pm1$.\newline
-- Particle identification is performed using the ratio $E/p$ of
energy deposition in the LKr calorimeter to
momentum measured by the DCHs. The vertex is required to be
composed of one $\pi$ candidate ($E/p<0.85$), and a pair of
oppositely charged $e^\pm$ candidates ($E/p>0.95$).

If several vertices satisfy the above conditions, the one with the
best vertex fit quality is considered. The $K^\pm\to\pi^\pm e^+e^-$
candidates are selected by applying the following criteria.\newline
-- $\pi^\pm e^+e^-$ momentum within the beam nominal range:
$54~{\rm GeV}/c<|\vec p_{\pi ee}|<66~{\rm GeV}/c$.\newline
-- $\pi^\pm e^+e^-$ transverse momentum with respect to
the beam trajectory (which is precisely measured using the the
concurrently acquired $K^\pm\to3\pi^\pm$ sample): $p_T^2<0.5\times
10^{-3}~({\rm GeV}/c)^2$.\newline
-- Kinematic suppression of the main background channel
$K^\pm\to\pi^\pm\pi^0_D$ (and other backgrounds induced by
$\pi^0_D$ and $\pi^0_{DD}\to4e^\pm$ decays) by requiring $z=(M_{ee}/M_K)^2>0.08$, which corresponds to $M_{ee}>140$~MeV/$c^2$, and leads
to a loss of $\sim 30\%$ of the signal sample.\newline
-- $\pi^\pm e^+e^-$ invariant mass:
$470~{\rm MeV}/c^2<M_{\pi ee}<505~{\rm MeV}/c^2$.
The lower limit corresponds to a $E_\gamma<23.1~{\rm MeV}$
cutoff for the energy of a single directly
undetectable soft IB photon.

For the $K^\pm\to\pi^\pm\pi^0_D$ normalisation mode candidates, a
presence of a LKr energy deposition cluster (photon candidate)
satisfying the following principal criteria is required.\newline
-- Reconstructed cluster energy $E>3$~GeV,
cluster time consistent with the vertex time, sufficient transverse
separations from track impact points at the LKr plane.\newline
-- $e^+e^-\gamma$ invariant mass compatible with a $\pi^0_D$
decay: $|M_{ee\gamma}-M_{\pi^0}|<10$~MeV/$c^2$.\newline
-- $\pi^\pm e^+e^-\gamma$
total and transverse momenta: same requirements as used for $K^\pm\to\pi^\pm e^+e^-$.\newline
-- $\pi^\pm e^+e^-\gamma$ invariant mass:
$475~{\rm MeV}/c^2<M_{\pi ee\gamma}<510~{\rm MeV}/c^2$.

\begin{figure}[tb]
\vspace{-2mm}
\begin{center}
{\resizebox*{0.42\textwidth}{!}{\includegraphics{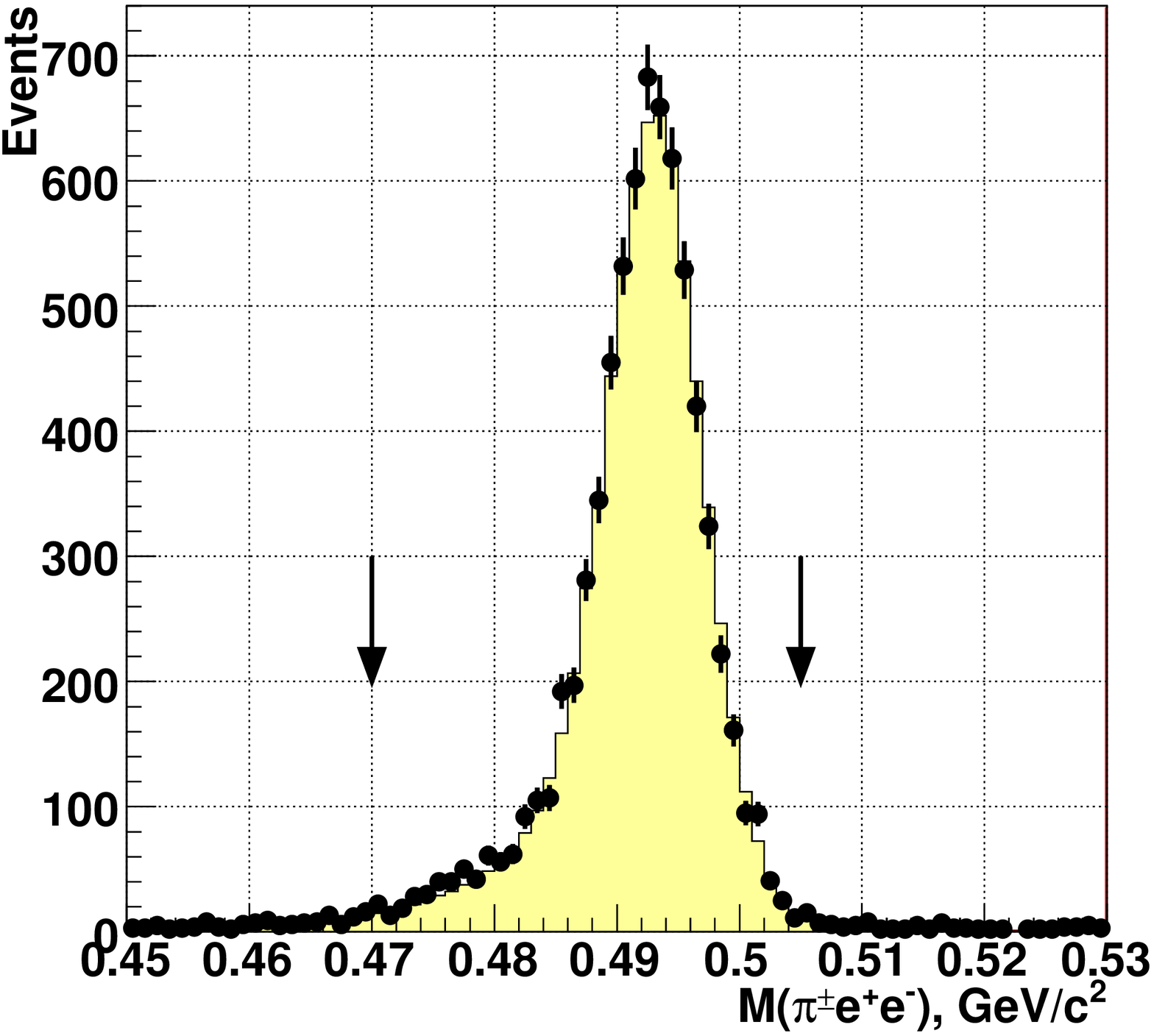}}}~~
{\resizebox*{0.4\textwidth}{!}{\includegraphics{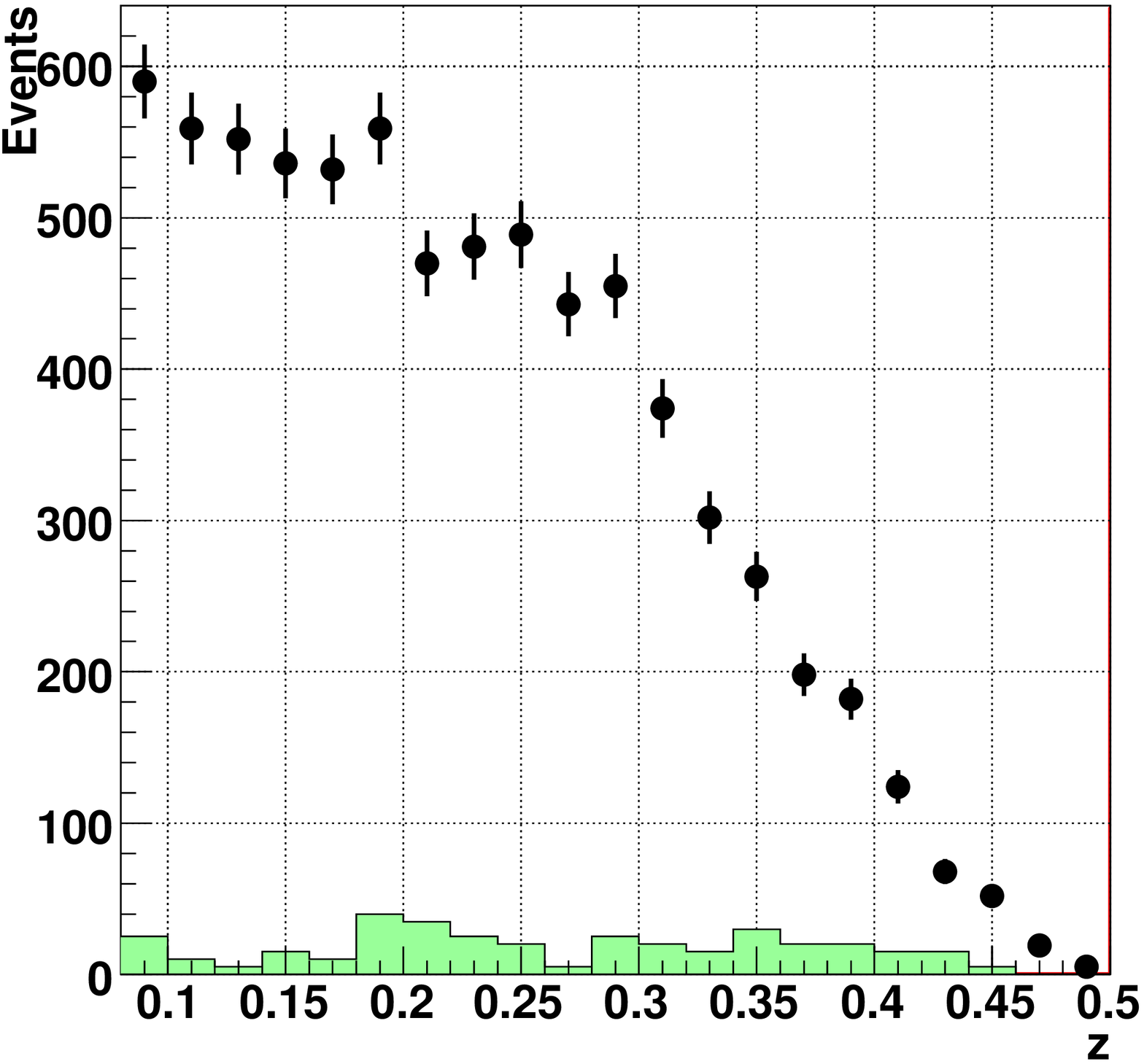}}}
\put(-207,145){\bf\large (a)} \put(-23,145){\bf\large (b)}
\end{center}
\vspace{-8mm} \caption{(a) Reconstructed spectrum of $\pi^\pm
e^+e^-$ invariant mass: data (dots) and MC simulation (filled area).
Note the description of the radiative mass tail by the PHOTOS
simulation. (b) $z$ spectrum of the selected
$K^\pm\to\pi^\pm e^+e^-$ candidates. Filled area: estimated background
multiplied by a factor of 5.} \label{fig:mk}
\end{figure}

The reconstructed $\pi^\pm e^+e^-$ invariant mass spectrum
is presented in Fig.~\ref{fig:mk}a. The number of
$K^\pm\to\pi^\pm e^+e^-$ candidates in the signal
region is $N_{\pi ee}=7253$, of which 4613 (2640) are $K^+$
($K^-$) candidates. The background sources are
$K^\pm\to\pi^\pm\pi^0_D$ and $K^\pm\to\pi^0_De^\pm\nu$ decays
with $\pi^0_D\to e^+e^-\gamma$ and $e^\pm/\pi^\pm$
misidentification, and
kaon decays with two or more $e^+e^-$ pairs in the final state
from $\pi^0_{D(D)}$ decays or external $\gamma$ conversions.
Background contamination is measured to be $(1.0\pm0.1)\%$ using the sum of spectra of the unphysical data
LFV $\pi^\mp e^\pm e^\pm$ and triple charge $\pi^\pm
e^\pm e^\pm$ candidates.

The number of $K^\pm\to\pi^\pm\pi^0_D$ candidates in the signal region is
$N_{2\pi}=1.212\times 10^7$. The only significant background source is the
$K^\pm\to\pi^0_D\mu^\pm\nu$ decay, with contamination in
the signal region estimated to be 0.15\% by MC simulation.

The decay is supposed to proceed through single virtual
photon exchange, resulting in a spectrum of the $z=(M_{ee}/M_K)^2$
kinematic variable sensitive to the form factor $W(z)$~\cite{da98}:
\begin{equation}
\frac{d\Gamma}{dz}=\frac{\alpha^2M_K}{12\pi(4\pi)^4}
\lambda^{3/2}(1,z,r_\pi^2)\sqrt{1-4\frac{r_e^2}{z}}
\left(1+2\frac{r_e^2}{z}\right)|W(z)|^2, \label{theory}
\end{equation}
where $r_e=m_e/M_K$, $r_\pi=m_\pi/M_K$, and
$\lambda(a,b,c)=a^2+b^2+c^2-2ab-2ac-2bc$.
The following parameterizations of the form factor $W(z)$ are
considered in the present analysis.
\begin{enumerate}
\item Linear: $W(z)=G_FM_K^2f_0(1+\delta z)$
with free normalisation and slope $(f_0,\delta)$. Decay rate and $z$
spectrum are sensitive to $|f_0|$, not to its sign.
\item Next-to-leading order ChPT~\cite{da98}:
$W(z)=G_FM_K^2(a_++b_+z)+W^{\pi\pi}(z)$ with free parameters
$(a_+,b_+)$ and an explicitly calculated pion loop term
$W^{\pi\pi}(z)$ given in~\cite{da98}.
\item Combined framework of ChPT and large-$N_c$ QCD~\cite{fr04}:
the form factor is parameterized as $W(z)\equiv W(\tilde{\rm
w},\beta,z)$ with free parameters $(\tilde{\rm w},\beta)$.
\item ChPT parameterization~\cite{du06} involving meson form
factors: $W(z)\equiv W(M_a,M_\rho,z)$. The resonance masses ($M_a$,
$M_\rho$) are treated as free parameters in the present analysis.
\end{enumerate}
The Coulomb factor is taken into account
following for instance~\cite{is08}. Radiative corrections
to $K^\pm\to\pi^\pm e^+e^-$ are evaluated with a
PHOTOS~\cite{photos} simulation of the
$K^\pm\to\pi^\pm\gamma^*\to\pi^\pm e^+e^-$ decay, and
cross-checked with a generalized computation for a multi-body
meson decay~\cite{is08}. They are crucial
for the extrapolation of the branching ratio from the limited
$M_{\pi ee}$ (equivalently, $E_\gamma$) signal region to the full
kinematic region: about 6\% of the total $K^\pm\to\pi^\pm e^+e^-(\gamma)$
decay rate fall outside the signal region $E_\gamma<23.1$~MeV.

The $z$ spectrum of the data events (in the
visible region $z>0.08$) presented in Fig.~\ref{fig:mk}b.
The values of $d\Gamma_{\pi ee}/dz$ in the centre of each $i$-bin of
$z$ are computed as
\begin{equation}
(d\Gamma_{\pi ee}/dz)_i = \frac{N_i-N^B_i}{N_{2\pi}}\cdot
\frac{A_{2\pi}(1-\varepsilon_{2\pi})}{A_i(1-\varepsilon_i)} \cdot
\frac{1}{\Delta z} \cdot \frac{\hbar}{\tau_K} \cdot {\rm
BR}(K^\pm\to\pi^\pm\pi^0)\cdot{\rm BR}(\pi^0_D). \label{dgdz}
\end{equation}
Here $N_i$ and $N^B_i$ are numbers of $K^\pm\to\pi^\pm
e^+e^-$ candidates and background events in the $i$-th bin,
$N_{2\pi}$ is the number of $K^\pm\to\pi^\pm\pi^0_D$ events (background subtracted), $A_i$ and $\varepsilon_i$ are geometrical
acceptance and trigger inefficiency in the $i$-th bin for the signal
sample (computed by MC simulation), $A_{2\pi}$ and
$\varepsilon_{2\pi}$ are those for $K^\pm\to\pi^\pm\pi^0_D$
events, $\Delta z$ is the bin width set to 0.02. The
external inputs are the kaon lifetime
$\tau_K$, and normalisation branching ratios ${\rm
BR}(K^\pm\to\pi^\pm\pi^0)$, ${\rm BR}(\pi^0_D)$.

\begin{figure}[t]
\begin{center}
{\resizebox*{0.4\textwidth}{!}{\includegraphics{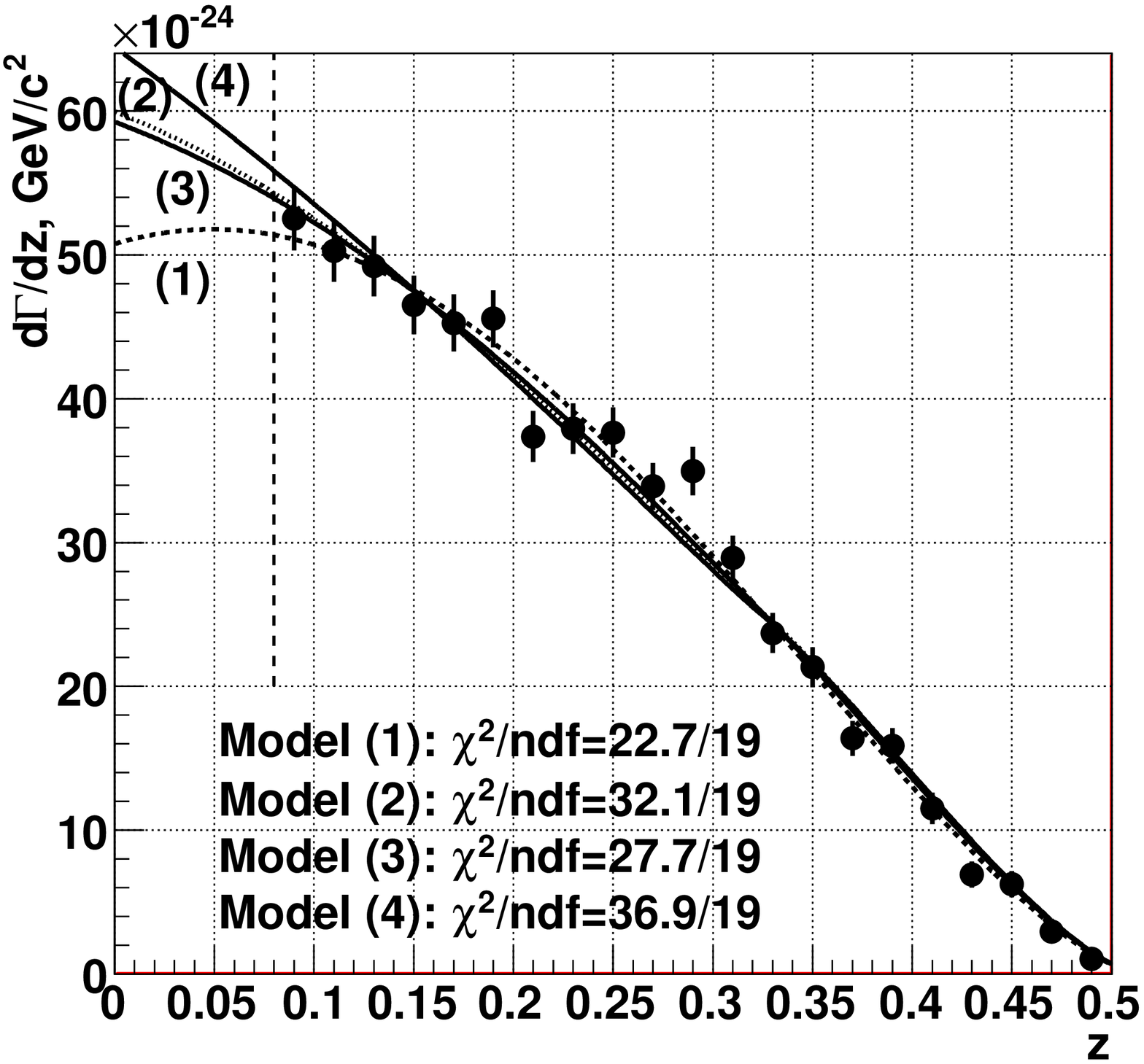}}}~~
{\resizebox*{0.4\textwidth}{!}{\includegraphics{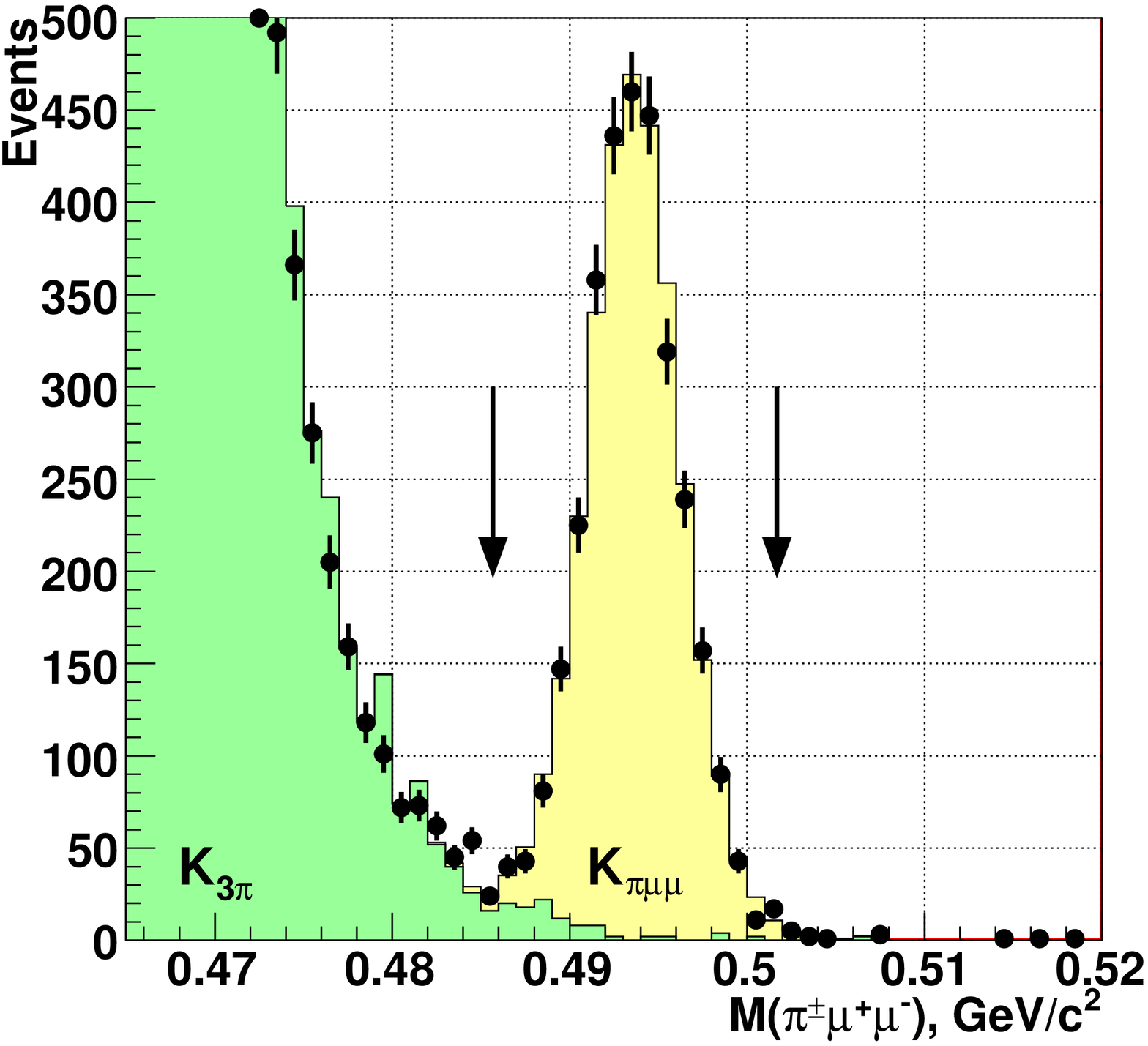}}}
\put(-205,140){\bf\large (a)} \put(-25,140){\bf\large (b)}
\end{center}
\vspace{-9mm} \caption{(a) $d\Gamma_{\pi ee}/dz$ (background
subtracted, corrected for trigger efficiency) and fit results
according to the four considered models. (b) Reconstructed spectrum of $\pi^\pm
\mu^+\mu^-$ invariant mass: data (dots), $K^\pm\to\pi^\pm
\mu^+\mu^-$ MC
simulation and $K^\pm\to 3\pi^\pm$ background estimate (filled areas).}
\label{fig:fit}
\end{figure}

The values of $d\Gamma_{\pi ee}/dz$ and results
of the fits to the four models are presented in
Fig.~\ref{fig:fit}a.
The model-independent branching ratio ${\rm BR_{mi}}$ in the
kinematic region $z>0.08$ is computed by integration of
$d\Gamma_{\pi ee}/dz$, and differs from each of the
model-dependent BRs computed in the same $z$ range by less than
$0.01\times10^{-7}$. The differences between model-dependent BRs
come from the region $z<0.08$, as seen in
Fig.~\ref{fig:fit}a.

Systematic uncertainties due to particle identification inefficiencies, imperfect MC description of the beamline, background subtraction, trigger inefficiency, radiative corrections, and fitting method are considered. The external uncertainties related to limited relative precision (2.7\%) of
${\rm BR}(\pi^0_D)$ are also taken into account.

\section{Results, discussion and prospects}

The measured model-independent ${\rm
BR_{mi}}(z>0.08)$, and the parameters of the considered
models are presented in
Table 2. The correlation coefficients and 68\% confidence contours for model parameters are shown in~\cite{piee}. Each of the considered models provides a reasonable fit to the data (as indicated in Fig.~\ref{fig:fit}a).
The data are insufficient to distinguish between
the models.
The measured form factor slope $\delta$
is in agreement with earlier measurements~\cite{al92,ap99,ma00}, and
disagrees to the meson dominance models~\cite{li99}
which predict lower slope values. The measured
$|f_0|$, $a_+$ and $b_+$ are in agreement with the previous
measurement~\cite{ap99}; $a_+$ is in agreement with a
theoretical prediction $a_+=-0.6^{+0.3}_{-0.6}$~\cite{br93}.
The measured $\tilde{\rm w}$,
$\beta$ are in fair agreement with an earlier measurement~\cite{ap99,fr04}.

\begin{table}[tb]
\label{tab:results}
\begin{center}
\caption{Model-independent ${\rm BR_{mi}}(z>0.08)$, and fit results
for the considered models.}
\begin{tabular}{rrrrrrrrrrrr}
\hline
${\rm BR_{mi}}\times10^7=\!\!\!$       &$\!2.28$  &$\!\!\pm\!\!$&$0.03_{\rm stat.}$ &$\!\!\pm\!\!$&$0.04_{\rm syst.}$ &$\!\!\pm\!\!$&$0.06_{\rm ext.}$ &$\!\!=\!\!$&$2.28$ &$\!\!\pm\!\!$&0.08\\
$|f_0|=\!\!\!$                         &$\!0.531$ &$\!\!\pm\!\!$&$0.012_{\rm stat.}$&$\!\!\pm\!\!$&$0.008_{\rm syst.}$&$\!\!\pm\!\!$&$0.007_{\rm ext.}$&$\!\!=\!\!$&$0.531$&$\!\!\pm\!\!$&0.016\\
$\delta=\!\!\!$                        &$\!2.32$  &$\!\!\pm\!\!$&$0.15_{\rm stat.}$ &$\!\!\pm\!\!$&$0.09_{\rm syst.}$ &             &                  &$\!\!=\!\!$&$2.32$ &$\!\!\pm\!\!$&0.18\\
$a_+=\!\!\!$                           &$\!-0.578$&$\!\!\pm\!\!$&$0.012_{\rm stat.}$&$\!\!\pm\!\!$&$0.008_{\rm syst.}$&$\!\!\pm\!\!$&$0.007_{\rm ext.}$&$\!\!=\!\!$&$-0.578$&$\!\!\pm\!\!$&0.016\\
$b_+=\!\!\!$                           &$\!-0.779$&$\!\!\pm\!\!$&$0.053_{\rm stat.}$&$\!\!\pm\!\!$&$0.036_{\rm syst.}$&$\!\!\pm\!\!$&$0.017_{\rm ext.}$&$\!\!=\!\!$&$-0.779$&$\!\!\pm\!\!$&0.066\\
$\tilde{\rm w}=\!\!\!$                 &$\!0.057$ &$\!\!\pm\!\!$&$0.005_{\rm stat.}$&$\!\!\pm\!\!$&$0.004_{\rm syst.}$&$\!\!\pm\!\!$&$0.001_{\rm ext.}$&$\!\!=\!\!$&$0.057$&$\!\!\pm\!\!$&0.007\\
$\beta=\!\!\!$                         &$\!3.45$  &$\!\!\pm\!\!$&$0.24_{\rm stat.}$ &$\!\!\pm\!\!$&$0.17_{\rm syst.}$ &$\!\!\pm\!\!$&$0.05_{\rm ext.}$ &$\!\!=\!\!$&$3.45$ &$\!\!\pm\!\!$&0.30\\
$M_a/{\rm GeV}/c^2)=\!\!\!$            &$\!0.974$ &$\!\!\pm\!\!$&$0.030_{\rm stat.}$&$\!\!\pm\!\!$&$0.019_{\rm syst.}$&$\!\!\pm\!\!$&$0.002_{\rm ext.}$&$\!\!=\!\!$&$0.974$&$\!\!\pm\!\!$&0.035\\
$M_\rho/({\rm GeV}/c^2)=\!\!\!$        &$\!0.716$ &$\!\!\pm\!\!$&$0.011_{\rm stat.}$&$\!\!\pm\!\!$&$0.007_{\rm syst.}$&$\!\!\pm\!\!$&$0.002_{\rm ext.}$&$\!\!=\!\!$&$0.716$&$\!\!\pm\!\!$&0.014\\
\hline
\end{tabular}
\end{center}
\vspace{-4mm}
\end{table}

The branching ratio in the full kinematic range, which includes a model-dependence uncertainty, is
${\rm BR}=(3.11\pm0.04_{\rm stat.}\pm0.05_{\rm syst.}\pm0.08_{\rm
ext.}\pm0.07_{\rm model})\times10^{-7}$;
it agrees to earlier measurements~\cite{bl75, al92, ap99}. The DCPV charge asymmetry of decay rates is measured for the first time:
$\Delta(K_{\pi ee}^\pm)=({\rm BR}^+-{\rm BR}^-)/({\rm BR}^++{\rm
BR}^-)=(-2.2\pm1.5_{\rm stat.}\pm 0.6_{\rm syst.})\times 10^{-2}$,
corresponding to an upper limit of $|\Delta(K_{\pi ee}^\pm)|<2.1\times10^{-2}$ at 90\% CL. However the achieved
precision is far from the SM expectation
$|\Delta(K_{\pi ee}^\pm)|\sim 10^{-5}$~\cite{da98} and even the SUSY
upper limit of $|\Delta(K_{\pi ee}^\pm)|\sim 10^{-3}$~\cite{me02}.

An analysis of the $K^\pm\to\pi^\pm\mu^+\mu^-$ decay based on the same data sample is in progress. A sample of $\sim 3100$ decay candidates with 3\% background from the $K^\pm\to3\pi^\pm$ is selected. The $\pi^\pm\mu^+\mu^-$ invariant mass spectrum is presented in Fig.~\ref{fig:fit}b. Unlike the $K^\pm\to\pi^\pm e^+e^-$ case, the full kinematic $z$ range is accessible, and effects of radiative corrections are suppressed. In addition to spectrum, rate and CPV measurements, the first measurement of the forward-backward asymmetry, which can be enhanced with respect to $K^\pm\to\pi^\pm e^+e^-$ in both SM and MSSM~\cite{ch03}, is performed.

\end{document}